\documentclass[12pt]{article}
\usepackage{slashed}
\usepackage{graphicx}

\def\endauthors{}
\def\authors#1\endauthors{#1}

\def\be{\begin{equation}}
\def\ee{\end{equation}}
\def\br{\begin{eqnarray}}
\def\er{\end{eqnarray}}
\def\brn{\begin{eqnarray*}}
\def\ern{\end{eqnarray*}}

\def\lbar{\mbox{$\lambda$\kern-0,450em \vrule width0,35em height1,252ex
depth-1,21ex \kern0,051em}}
\def\dbar{\mbox{d\kern-0,347em \vrule width0,3em height1,252ex depth-1,21ex
\kern0,051em}}
\def\Dbar{\mbox{D\kern-0,735em \vrule width0,3em height0,86ex depth-0,81ex
\kern0,40em}}

\def\vb {{\bf v}}

\def\mbpi{\mbox{\boldmath$\pi$}}

\def\mbl{\mbox{\boldmath$\lambda$}}

\def\g {{\gamma}}

\def\N {{{\cal N}}}

\def\S {{{\cal S}}}

\def\ba#1{\begin{array}{#1}}
\def\ea{\end{array}}

\def\bc{binomial coefficients }

\def\be{\begin{equation}}
\def\ee{\end{equation}}
\def\br{\begin{eqnarray}}
\def\er{\end{eqnarray}}
\def\brn{\begin{eqnarray*}}
\def\ern{\end{eqnarray*}}
\def\bit{\begin{itemize}}
\def\eit{\end{itemize}}
\def\bnu{\begin{enumerate}}
\def\enu{\end{enumerate}}

\def\={{\simeq}}

\def\nn{\nonumber }

\def\2q{{{\{}2{\}}_q}}
\def\3q{{{\{}3{\}}_q}}
\def\mbk{\mbox{\boldmath$k$}}
\def\mbpi{\mbox{\boldmath$\pi$}}
\def\mbN{\mbox{\boldmath$N$}}
\def\mbl{\mbox{\boldmath$l$}}
\def\mbnu{\mbox{\boldmath$\nu$}}

\def\be{\begin{equation}}
\def\ee{\end{equation}}
\def\br{\begin{eqnarray}}
\def\er{\end{eqnarray}}
\def\bc{\begin{center}}
\def\ec{\end{center}}

\def\N3{{1\over (2\pi)^3}}

\def\g5{\gamma_5}

\def\endauthors{}
\def\authors#1\endauthors{#1}

\def\g {{\gamma}}

\begin{document}

\title{2p2h effects on the weak pion production cross section}
\author{A. Mariano$^\dag$ and C. Barbero$^\dag$}
\date{$^\dag$Departamento de F\'isica, Facultad de Ciencias Exactas, Universidad Nacional de La Plata - IFLP (Conicet), C.C. 67, La Plata, Argentina}

\maketitle
\begin{abstract}
The $\nu_l n \rightarrow l^- p$ QE reaction  on the A-target is used
as a signal event or/and to reconstruct the neutrino energy,  using two-body kinematics. Competition of another processes could lead to misidentification of the arriving neutrinos, being important  the fake events coming from the CC1$\pi$ background. A precise knowledge of cross sections is a prerequisite in order to
make simulations  in event generators to substract the fake ones  from the  QE countings, and in this contribution we analyze the different nuclear effects on the CC1$\pi$ channel. Our calculations also can be extended for the NC case.
\end{abstract}

$\nu_\mu \rightarrow \nu_x$ disappearance experiments
uses $\nu_\mu n \rightarrow \mu^- p$ CCQE to detect neutrinos and
reconstruct its energy. $E_\nu$ determination could be wrong for a fraction of  CC1$\pi^+$ background events (20\%)  $ \nu_\mu p \rightarrow \mu^- p \pi^+$,  that can mimic a CCQE one if the pion is absorbed and/or not detected.
 These processes take place into the target nucleus and nuclear effects as smearing (S) of the reconstructed energy by the momentum distribution ($n_A$) of the target binding (B) nucleons, should be taken into account.
In addition final state interactions (FSI) of the emerging hadrons generate energy lost,change of direction,charge transfer or multiple nucleon knock out(np-nh). Finally meson exchange currents (MEC) processes lead to additional contributions to one-body current generated. In what follows we concentrate on the 2p2h+1$\pi$ contributions to the pion  production cross section, and compare with the 1p1h+1$\pi$ one already analyzed previously \cite{Lalakulich2012}.
The 2p2h+1$\pi$ amplitude is depicted in Figure 1
and  the corresponding differential cross section reads (${\cal N}^2(\mbk)\equiv { f\over (2\pi)^3E(\mbk)}, f=1/2(
M)$ for bosons(fermions))

\begin{figure}
  \begin{center}
  \vspace{-1.0cm}
    \includegraphics[width=9.cm]{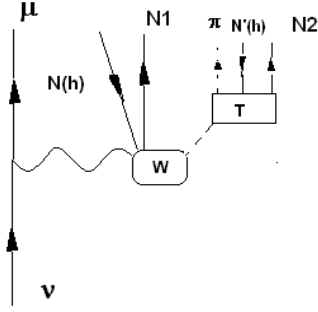}
  \end{center}
  \vspace{-8cm}
  \caption{\small Amplitude for the $\nu N \rightarrow l N' N_1 N_2  \pi$ ($2p2h + 1\pi$)  process.}
  \label{oneloop}
\end{figure}
\br
&&d\sigma_{\nu A}^{2p2h+1\pi} =  {n_A(\mbN)(1-n_A(\mbN_1))\over |\vb_\nu - \vb_A|2 E(\mbnu)}
{1 \over 2}\sum_{m_\nu,m_N,m_l,m_{N_1}} \int {d^4\pi'\over (2\pi)^4 }  \nn\\
&\times&\left [  \S \sum_{m_{N_2} m_N'}|T(N_2 m_{N_2},\pi,N'm_{N'},\pi')|^2{n_A(\mbN')(1-n_A(\mbN_2)) \over (\pi'^2 - M_\pi^2)^2}\right.\nn\\
&\times&\left. (2\pi)^4 \delta^4(N_2+\pi-\pi'-N')d^3N_2 {\cal N}^2(\mbN_2)d^3N' {\cal N}(\mbN')d^3\pi{\cal N}^2(\mbpi) \frac{}{}\right]\nn\\
&\times&|W^\mu(N_1 m_{N_1},l m_l, \pi', N m_N,\nu m_\nu) J^l_\mu(l m_l, \nu m_\nu)|^2 \nn\\
&\times&(2\pi)^4 \delta^4(N_1+\pi'+l-\nu-N)~ d^3 l {\cal N}^2(\mbl)~ d^3 N_1 {\cal N}^2(\mbN_1)
d^3N {\cal N}(\mbN),\nn
\er
being $N,\pi,l,\nu \equiv (E(\mbN,\mbpi,\mbl,\mbnu), \ \mbN,\mbpi,\mbl,\mbnu)$, $E(\mbk)=\sqrt{\mbk^2+M^2}$, $m\equiv$ spin, $\S$ symmetrization factor, and T $\pi' N'\rightarrow \pi N_2$ rescattering is simplified replacing
\br
&&{\bf[ \cdot \cdot \cdot]} \Rightarrow {1 \over (\pi'^2 - M_\pi^2 - \Pi(\pi'))} \Rightarrow 2\pi \delta(\pi'^2 - M_\pi^2 - \Re(\Pi(\pi'))),\nn
\er
as shown in Figure 2. We work out this self-energy $\Pi(\pi')$ in the $\Delta-h$ approach, taking into account that the $\Delta$ width will account the
final pion-nucleon additional state through the absorptive evaluation of the pion-nucleon $\Delta$-self energy.
Figure 3 shows results for the total CC$1\pi^+$ cross section and the
\nolinebreak gradual effect of the B, S and FSI within the 1p1h+1${\pi}$ configuration space. Also \nolinebreak the
results for B+S+FSI in the 2p2h+1${\pi}$ one and full 1p1h+2p2h+1${\pi}$ \nolinebreak one are \nolinebreak shown. They are also shown for the differential cross section in Figure 4.\nolinebreak
B effects are considered within the Relativistic Hartree approximation (RHA) of QHD I \cite{Serot86}, for N and $\Delta$ using universal couplings.
$n_A$ is obtained from a perturbative approach in nuclear matter within a 2p2h+4p4h configuration space\cite{Mariano96}. FSI on nucleons is taken (Toy model !) through the RHA effective fields also for final N, while for pions we use the Eikonal approach\cite{fsi}.

\begin{figure}[h!]
  \begin{center}
  \vspace{-0.8cm}
    \includegraphics[width=13.5cm]{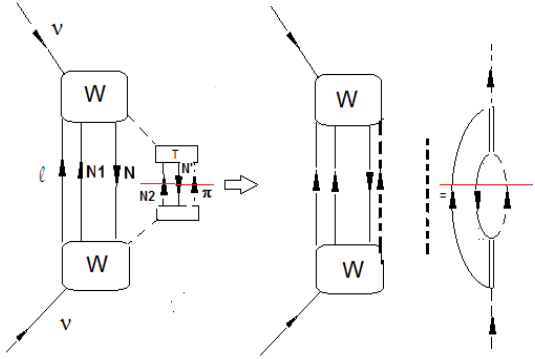}
  \end{center}
  \vspace{-12.5cm}
  \caption{\small Simplification to calculate $\pi' N'(h)$ rescattering.}
  \label{oneloop}
\end{figure}
\vspace{-0.35cm}

\begin{figure}[h!]
 \begin{center}
  \vspace{-1.cm}
    \includegraphics[width=10.8cm]{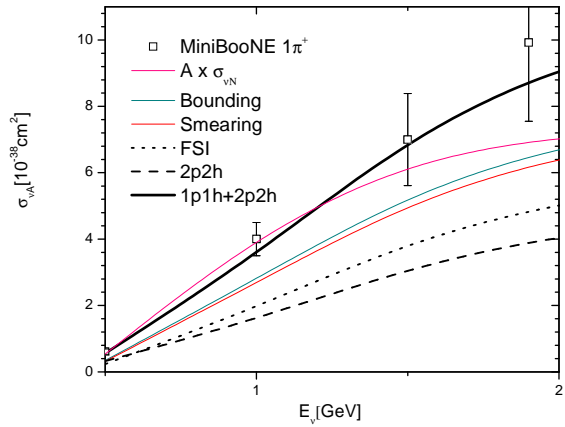}
 \end{center}
  \vspace{-7.3cm}
  \caption{\small Total 1$\pi^+$ cross section compared with MiniBooNE data (see\cite{Lalakulich2012}).}
  \label{oneloop}
\end{figure}
\begin{figure}[h!]
  \vspace{-1.cm}
\begin{center}
 \includegraphics[width=10.cm]{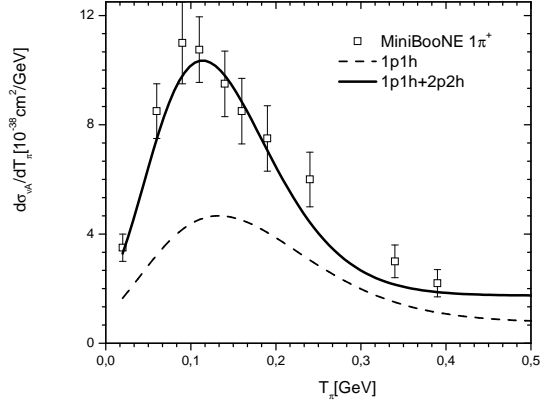}
\end{center}
 \vspace{-7.5cm}
   \caption{\small 1$\pi^+$ production differential cross section.}
  \label{oneloop}
\end{figure}
\vspace{-0.30cm}

We conclude that 2p2h contribution is important and comparable to the 1p1h one.
In the $\pi^0$ channel results are not so good in reproducing the data and nonresonant contributions and charge exchange terms should be included in the rescattering amplitude. Finally, MEC should be included at the same time that 2p2h contributions in order to have a more real estimation.

 \vspace{0.35cm}

Acknowledgments:
The assistance of A.M. to the Nuint 2012 was supported partially by the CONICET under the PIP 0349.

\end{document}